\newcommand{\xmax}{\ensuremath{X_{\rm max}}}

\documentclass[
    ,final            % use final for the camera ready runs
  ]
  {aipproc}

\layoutstyle{6x9}

\begin{document}

\title{NICHE: The Non-Imaging CHErenkov Array}


\author{Douglas Bergman}{
  address={Dept. of Physics and Astronomy, University of Utah, Salt Lake City, UT 84112 USA}
}

\author{John Krizmanic}{
  address={Universities Space Research Association, Columbia, MD 21044 USA}
}

\begin{abstract}
The accurate measurement of the Cosmic Ray (CR) nuclear composition around and above the Knee ($\sim 10^{15.5}$ eV) has been difficult
% is crucial in understanding the evolution of the high-energy end of the galactic Cosmic Ray spectrum. %However, the current understanding is ambiguous 
due to uncertainties inherent to the measurement techniques and/or dependence on hadronic Monte Carlo simulation models required to interpret the data.   Measurement of the Cherenkov air shower signal, calibrated with air fluorescence measurements, offers a methodology to provide an accurate measurement of the nuclear composition evolution over a large energy range.
NICHE will use an array of widely-spaced, non-imaging Cherenkov counters to measure the amplitude and time-spread of the air shower Cherenkov signal to extract CR nuclear composition measurements and to cross-calibrate the Cherenkov energy and composition measurements with TA/TALE fluorescence and surface detector measurements. 

\end{abstract}

\maketitle

%%%%%%%%%%%%%%%%%%%%%%%%%%%%%%%%%%%%%%%%%%%%
%% MAINMATTER
%%%%%%%%%%%%%%%%%%%%%%%%%%%%%%%%%%%%%%%%%%%%

\section{Scientific Motivation}

Measurement of the changing nuclear composition of High-Energy Cosmic Rays (HECRs) at energies about and above the Knee, $>\! 10^{15}$ eV, provides a unique tool for understanding the evolution of the high-energy end of the galactic CR spectrum. This will in turn provide a firm foundation for understanding the composition and spectrum of extragalactic UHECRs, which overtake galactic HECRs above the 2nd Knee $\sim\! 10^{17.5}$ eV. The current understanding of the average composition in the Knee region is muddled, as illustrated in Figure 1, which uses
 data from a recent listing of experimental measurements\cite{Blumer2009}.
Different measurement techniques, sometimes performed within the same experiment, lead to qualitatively different assessments of the energy dependence of the average composition. The situation is further complicated in that interpretation of the experimental results may rely heavily on Monte Carlo studies\cite{Kampert2012}, thus subject to systematic effects due to the assumed hadronic simulation model. This ambiguity makes it difficult to confirm the widely held hypothesis that the Knee is the result of a Peters cycle\cite{Peters1960}, with rigidity-dependent cutoffs of the various galactic CR nuclear components.

One way to minimize this state of uncertainty is to perform a coordinated set of experiments to observe the composition of CRs over many orders-of-magnitude, using techniques that are directly sensitive to composition-dependent observables. One sensitive observable for measuring CR composition in air showers is the depth of shower maximum, \xmax, which is directly accessible to both Fluorescence Telescopes (FTs) and Cherenkov detectors. Traditionally, non-imaging Cherenkov detectors and FTs have not been used to overlap in energy. However, by using both the lateral distribution of Cherenkov photons and the time-domain structure of their arrival times,  the spacing of Cherenkov counters can be sufficiently increased to overlap with lower-energy FTs such as TALE.

\begin{figure}
  \includegraphics[height=.4\textheight]{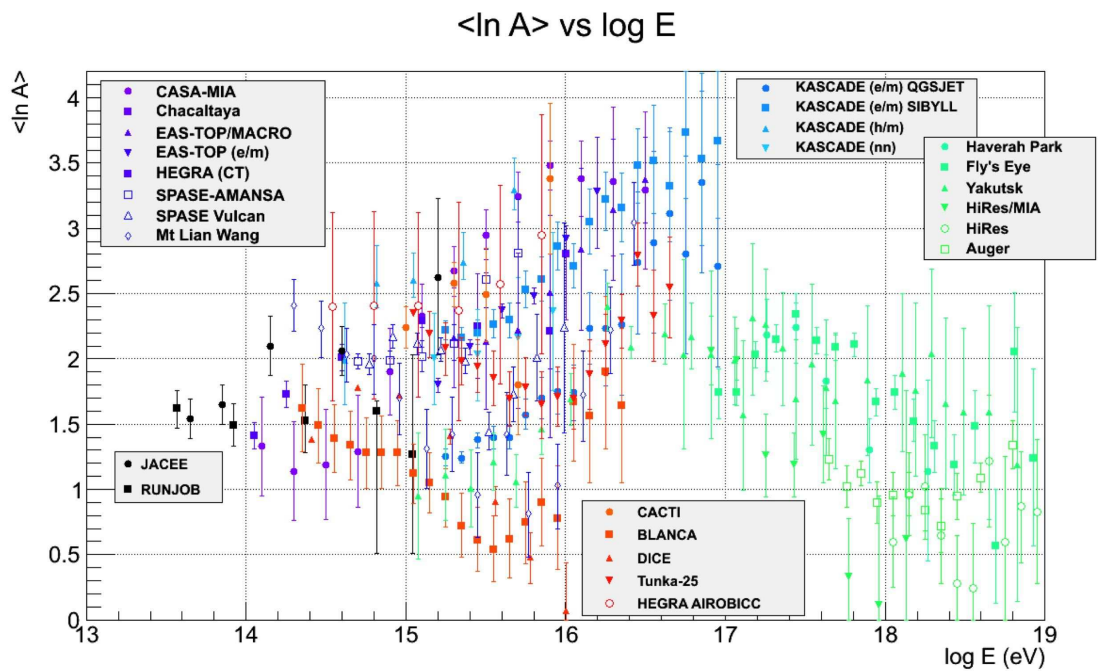}
 \caption{A compilation of CR composition measurements, as listed in \cite{Blumer2009}, given as average ln(A) where A is the atomic number, as a function of energy for a number of experiments.}
\end{figure}

The Telescope Array Low Energy (TALE) extension is now being deployed, with a low-energy threshold near $10^{16.5}$ eV. TALE is a hybrid detector combining both a Surface Detector (SD) array and Fluorescence Telescopes. A wide-spaced, Non-Imaging CHErenkov (NICHE) array built and deployed co-sited with TALE would allow the cross-calibration of the time-domain, non-imaging Cherenkov technique with direct \xmax\, measurements from the TALE FTs. The NICHE array will also provide core position and shower angle measurements comparable in performance to the TALE SD array. In addition, the Cherenkov array will extend the energy range of the TA/TALE/NICHE ensemble down to the CR Knee and potentially below.

\section{The Non-Imaging Cherenkov Technique}

Energetic electrons in an EAS produce Cherenkov radiation if they move faster than the speed-of-light in the local medium. The index of refraction in the atmosphere increases with depth, leading to Cherenkov cones from altitudes 8-20 km to overlap in a ring of radius 120-140 m at the ground, shown schematically in the left plot in Figure 2. The interior of the ring is filled by the portion of the air shower at lowest altitudes. Showers that develop deeper in the atmosphere will thus have a larger interior-to-ring ratio. Electron transverse momentum somewhat smears the Cherenkov ring on the ground, but the inside-to-outside ratio remains composition, e.g. depth, dependent.

At a given point on the ground, a counter will observe two components of Cherenkov light: one from the bulk of the shower, where some fraction of the transverse electrons are pointing their Cherenkov cones at the counter; and another part due to the small portion of the shower core where its Cherenkov cone intercepts the counter. Photons in the former component arrive over a long time span because the measurement samples a large portion of the developing air shower, while the latter component is narrow in time. When there is no dominant core component, the FWHM can be quite wide. The FWHM in time thus depends on shower development: deeper showers will have more Cherenkov light coming late, effectively sampling the air shower over a long path length. The right plot in Figures 2 illustrate these effects for $10^{16.5}$ eV vertical showers as a function of distance from the shower core: showers with deeper \xmax\, have a wider temporal FWHM for a given distance.

Various experiments have employed the non-imaging Cherenkov technique (AIROBICC\cite{AIROBICC1995}, BLANCA\cite{BLANCA2001}, CACTI\cite{CACTI2003}, and Tunka\cite{Tunka2011}) using the Cherenkov Light Distribution (CLD) to measure the CR spectrum, while one experiment (BASJE\cite{BASJE2009}) has employed the Cherenkov Time Domain technique, albeit in a small array. The innovation of NICHE is to combine these two techniques to construct an array of sufficient area to have significant overlap with TA/TALE air fluorescence measurements, leading to a cross-calibration of the energy scale. 

\begin{figure}
  \includegraphics[height=.3\textheight]{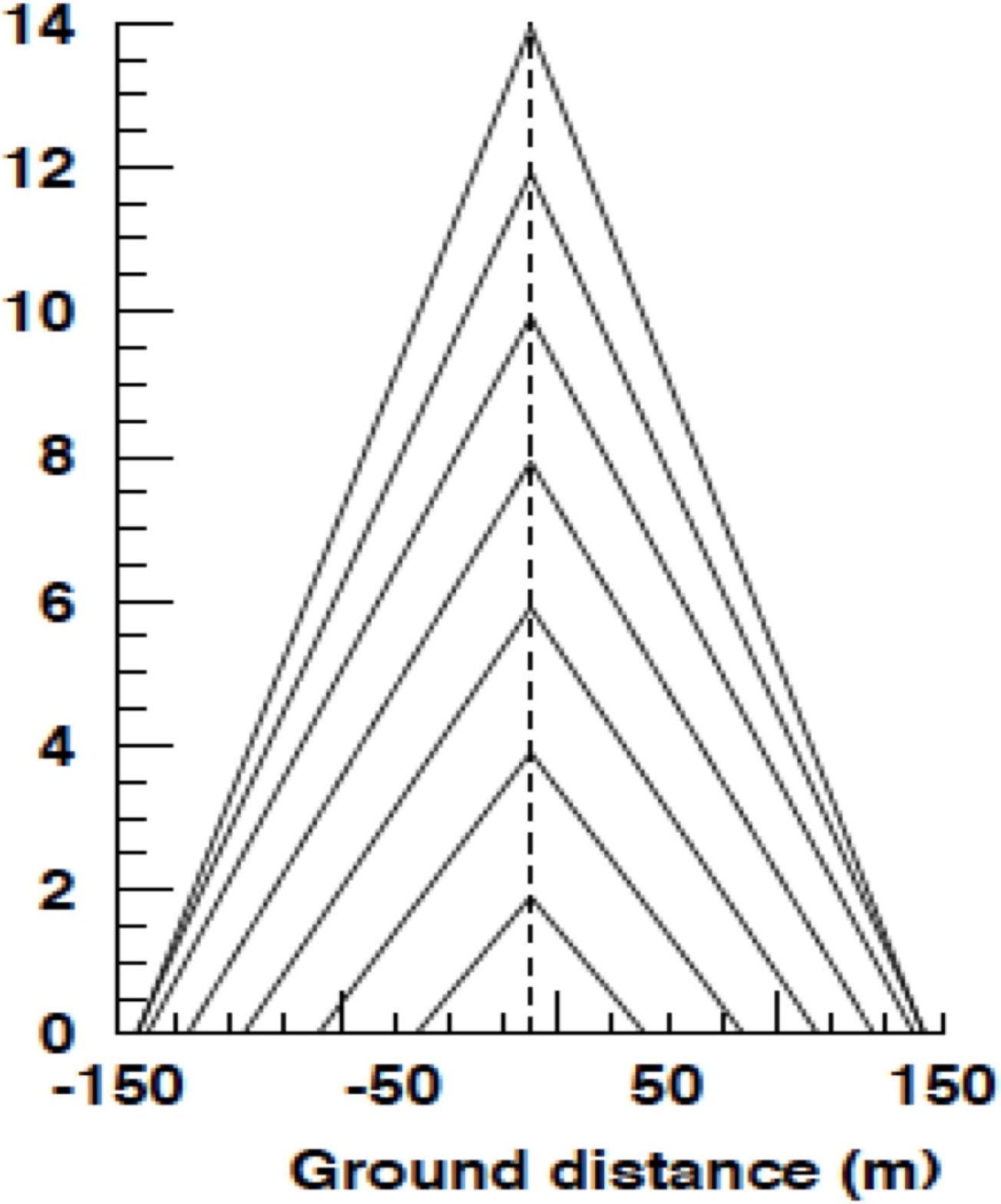}
\hspace{0.2cm}
 \includegraphics[height=.3\textheight]{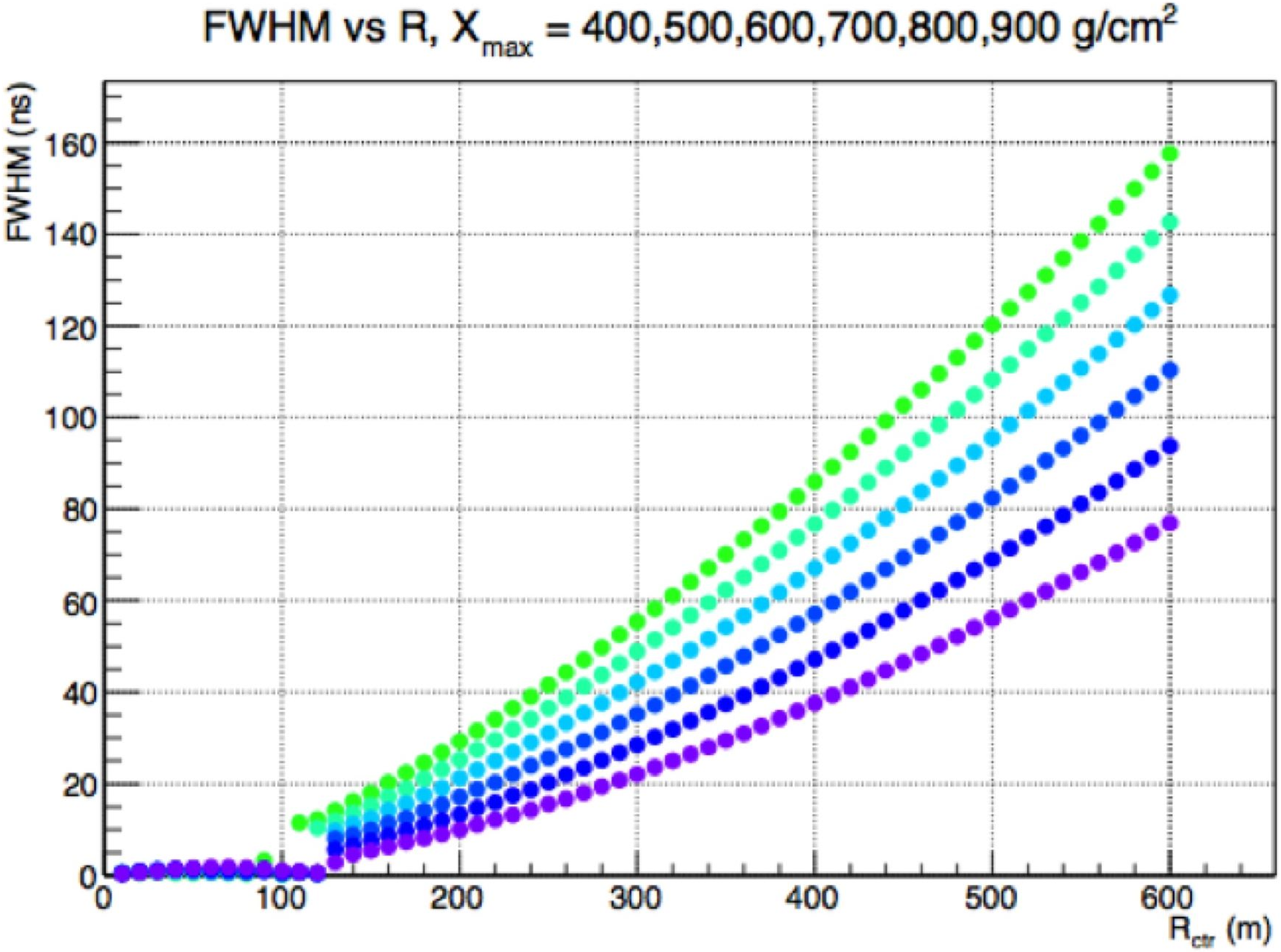}
  \caption{Left: Diagram of air shower Cherenkov cones as a function of emission height (km) in the absence of particle angular spread. Right: The simulated FWHM of the arrival time distribution as a function of distance to the core for different \xmax\,  (vertical showers, $10^{16.5}$ eV). From low to high, \xmax= 400 g/cm$^2$ to \xmax= 900 g/cm$^2$ in steps of 100 g/cm$^2$. The FWHM in discontinuous at around 120 m.}
\end{figure}

\section{NICHE Counter Design}

NICHE's individual Cherenkov counters are based on the BLANCA counter design\cite{BLANCA2001}: each uses a 3$^{\prime \prime}$ PMT with Winston cone. The cone acceptance angle has been increased to 35$^\circ$ for NICHE in order to increase acceptance.  Key improvements are to pair the PMT with a 200 MHz FADC DAQ system and to use GPS for time tagging. The system is designed to be powered remotely through battery and solar cells, and to communicate remotely through a radio-based WLAN system. Level two triggers will be provided by inter-counter IR communication. A schematic of a NICHE counter and electronics is shown in the left plot in Figure 3.

\begin{figure}
  \includegraphics[height=.32\textheight]{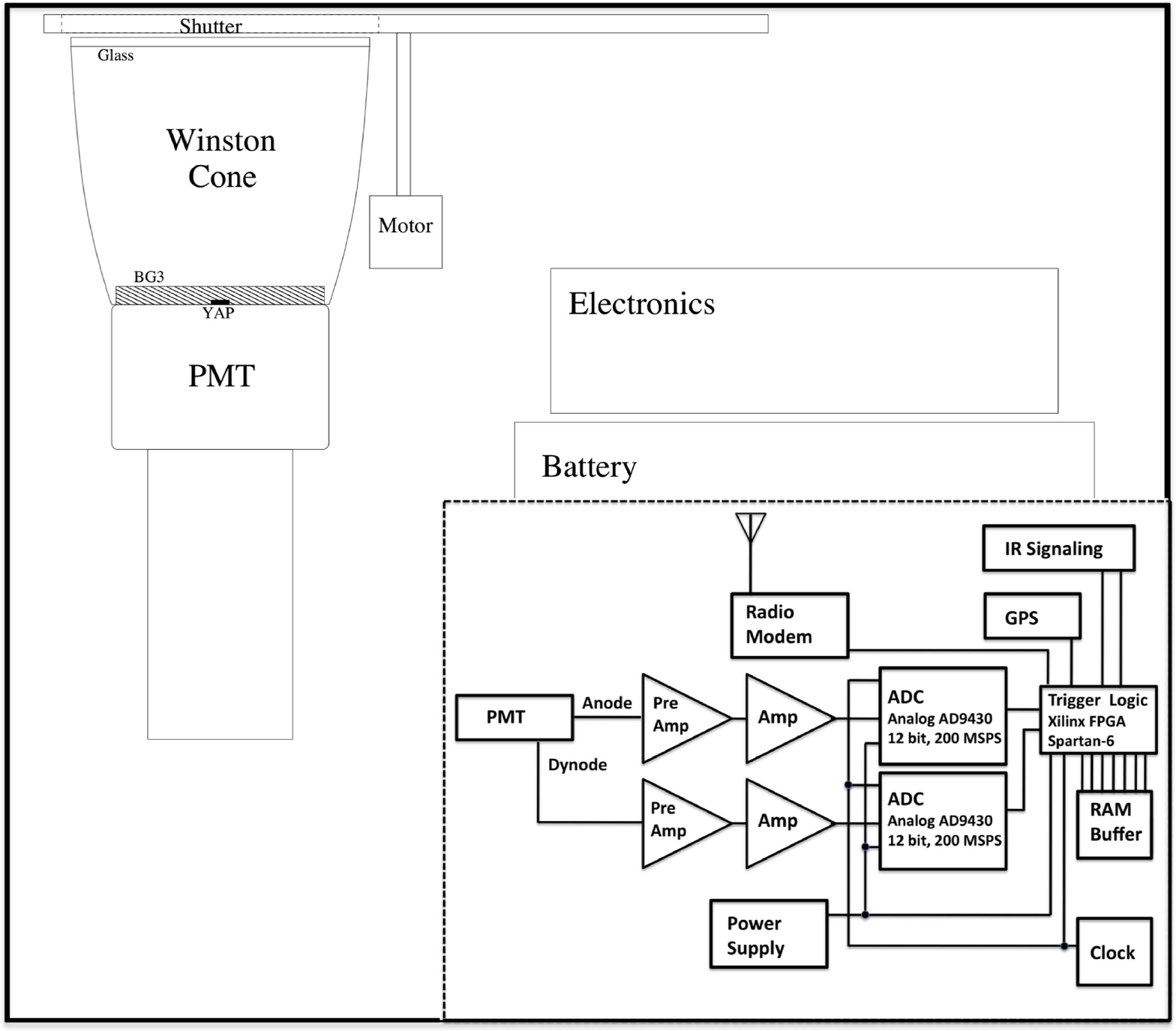}
\hspace{0.5cm}
 \includegraphics[height=.32\textheight]{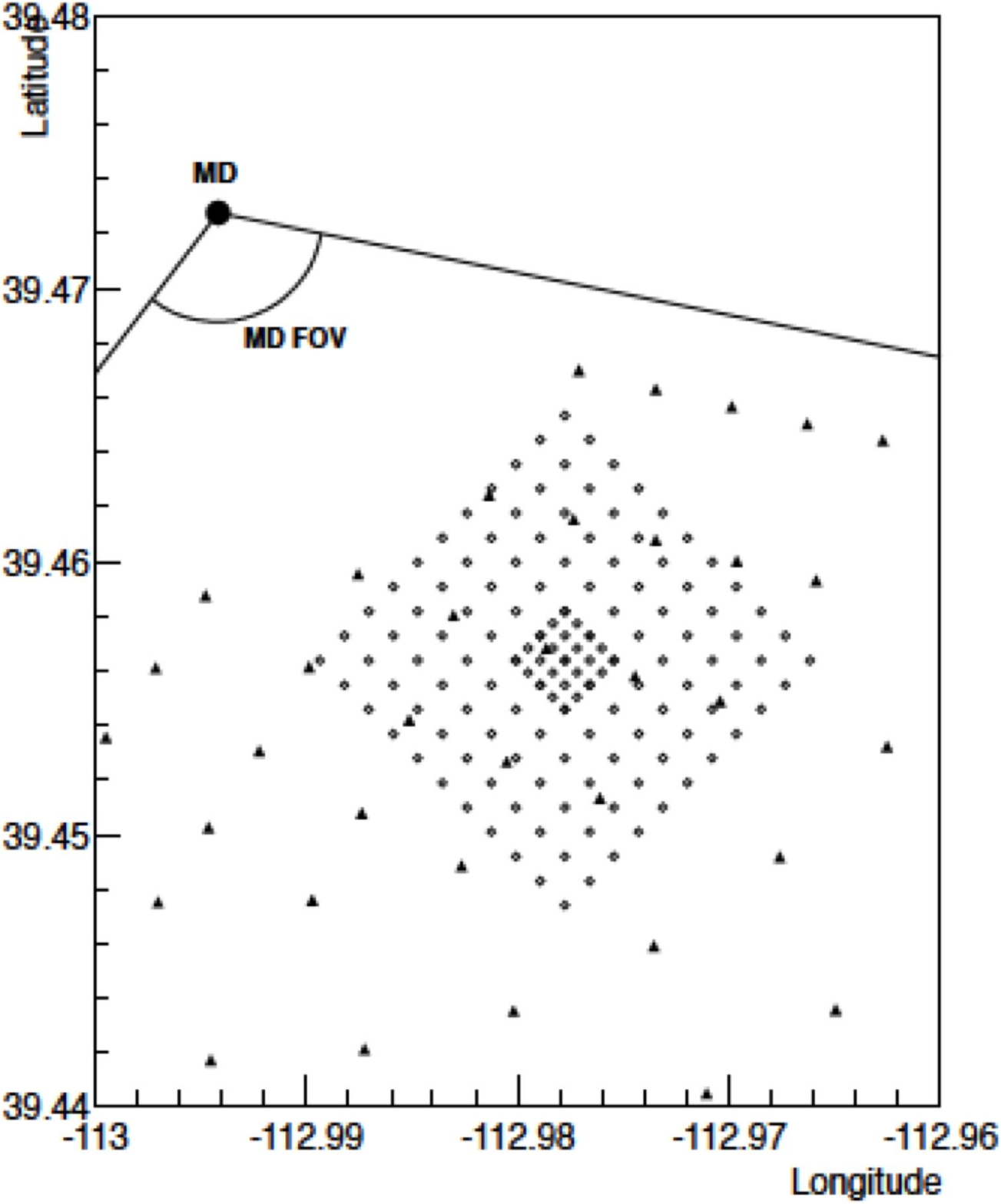}
  \caption{Left: Schematic of the side view of single NICHE detector. The Inset is a block diagram of the data acquisition electronics for the detector. Right: The expected layout of NICHE (small open circles) within the TA TALE extension.  The Middle Drum FT station is a large solid circle, with its field-of-view indicated. The TALE SD detectors are solid triangles and squares are counters in the TA SD array.}
\end{figure}

\section{The NICHE Array \& TALE}

The NICHE array is designed to have enough aperture above $10^{17}$ eV to have a significant overlap with the TALE fluorescence and surface detector measurements.  The NICHE instrumented area is 2 km$^2$ and when combined with an angular acceptance of zenith angles to 35$^\circ$ will result in collecting 1000 events per year above $10^{17}$ eV. For an $11 \times 11$ array, this results in an inner-counter separation of 140 m. This counter separation ensures that at least one counter, and usually more, provides measurements of the intensity and time-width of the Cherenkov light within the 120-m ring, while there are significantly more measurements outside the ring.

The NICHE array will be placed within the field-of-view of the TALE FT, as shown in Figure 3, and also be interspersed with the TALE SDs. The NICHE array must be close to the FT detector to provide the largest possible overlap, in terms of energy measurement, between the two disparate systems. The TALE detectors will add considerably to the aperture for time-domain measurements of NICHE as they will provide independent shower core position measurements for air showers that land outside of NICHE.

To improve detection and reconstruction at low energies, where NICHE will act as a stand-alone detector, a small part of the array will have 70-m spacing. Low-energy events will trigger counters within a smaller radius, thus requiring the smaller spacing. The smaller aperture is compensated by a much larger CR flux at these lower energies.

\section{Simulated Performance}

\begin{figure}
 \includegraphics[height=.24\textheight]{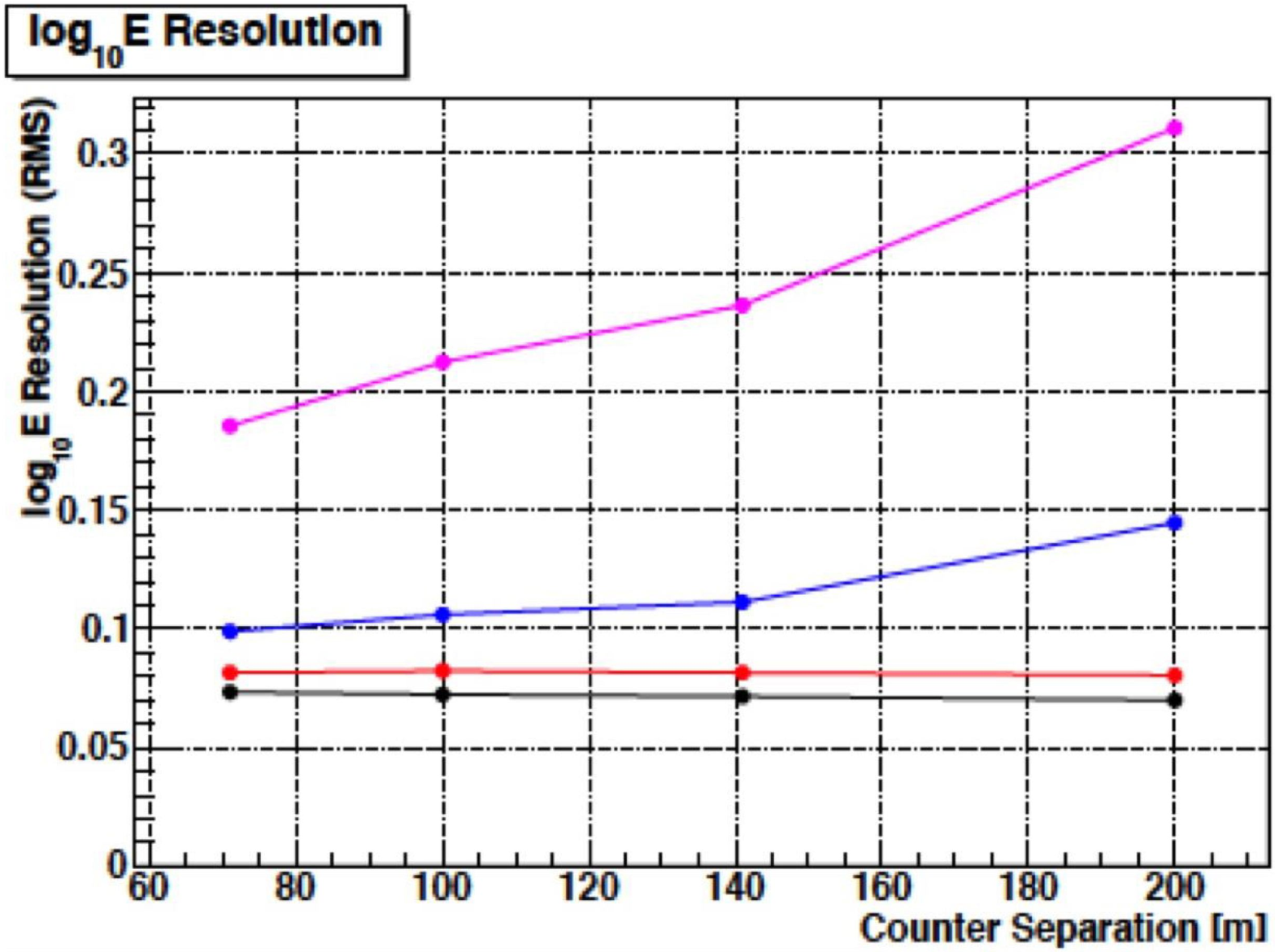}
\hspace{0.2cm}
 \includegraphics[height=.24\textheight]{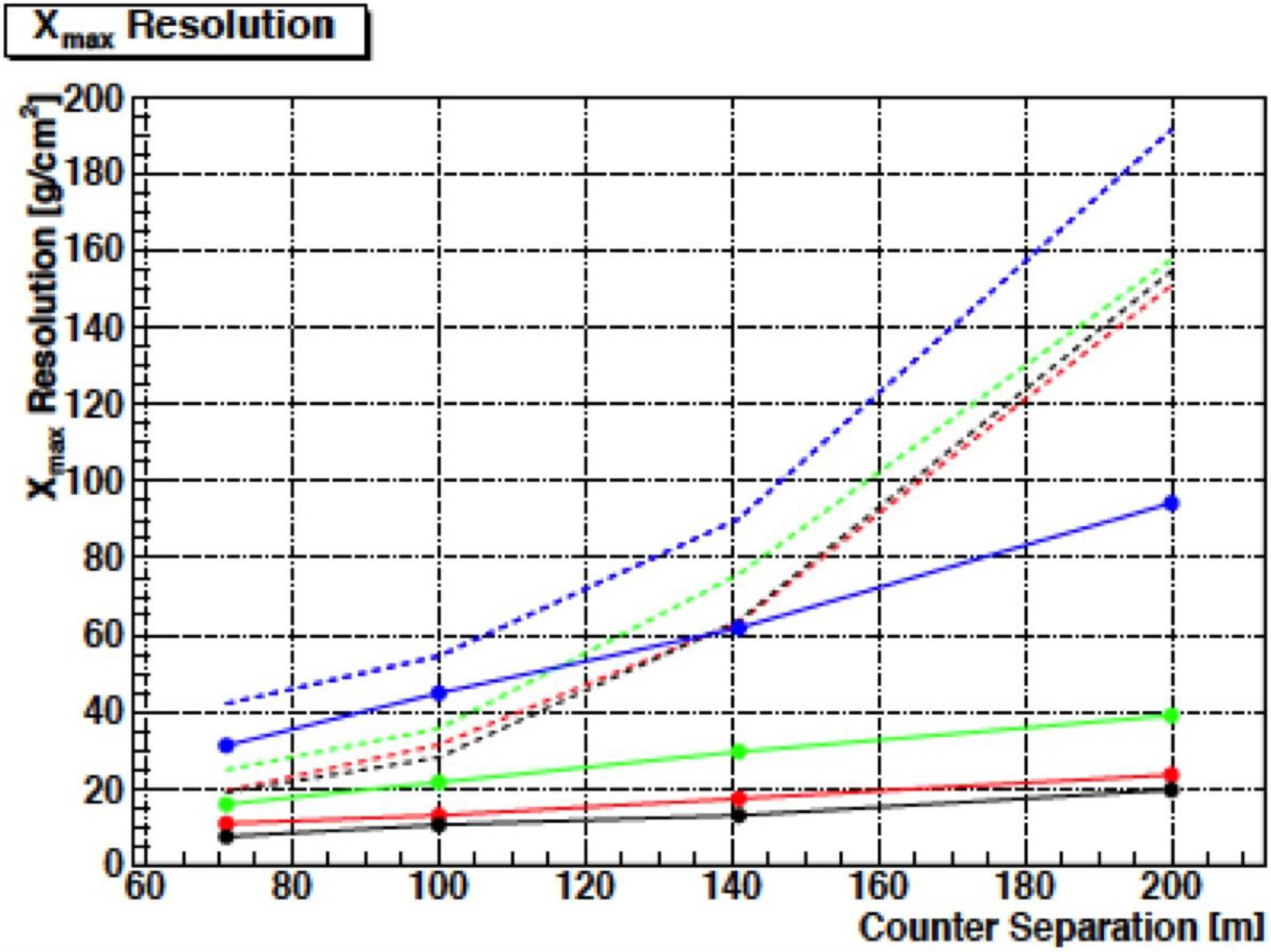}
  \caption{Left: The expected energy resolution as function of counter separation for air showers with log$_{\rm 10}$(E/eV) equal to 15.5, 16, 17, and 17.5 (from high to low) using 3-inch PMTs with 5-ns rise-time. The resolution is given as the RMS of log$_{\rm 10}$(Erec/Egen).
Right: The expected \xmax\, resolution as function of counter separation for showers with log$_{\rm 10}$(E/eV) equal to 16, 16.5, 17, and 17.5 (from high to low) using 3-inch PMTs with 5-ns rise-time. The resolution using the Cherenkov Lateral Distribution is shown using dashed lines, while the resolution using time-width measurements is shown using solid lines.} 
\end{figure}

We have simulated the NICHE array in CORSIKA assuming 140-m counter separation. The PMT response was modeled using a 3rd order Butterworth filter ($t^2 e^{-t}$ in the time domain) with a time-scale of 5 ns. The time-width at a distance of 320 m from the shower core was used to reconstruct the distance to shower max for a variety of both proton and iron showers with an energy of $10^{16.5}$ eV. 

Figure 4 details the energy and \xmax\, resolution as a function of energy based upon Corsika simulation studies. The expected energy resolution of NICHE measurements is determined to be approximately 15\% above $10^{17}$ eV, slowly worsening as the incident energy decreases to $10^{16}$ eV before rapidly increasing to 50\% at $10^{15.5}$ eV. Note that these simulated results did not incorporate the measurements of the 70-m spacing in-fill array or Winston cones, both of which will improve performance. The \xmax\, resolution is 20 g/cm$^2$ above $10^{17}$ eV and 30 g/cm$^2$ at $10^{16.5}$ eV. Note that the time-domain analysis gives considerably better composition measurements at 140-m spacing than the inner-outer ratio measurements of the CLD.

NICHE will be able to measure the cosmic ray composition up to an energy of $10^{17}$ eV within a few years. The example results shown in Figure 5 demonstrate that a two or three component model could be excluded in the case that the actual data was generated as four-component mixture of H, He, CNO, and Fe in the ratio 1:1:1:7.

\begin{figure}
  \includegraphics[height=.43\textheight]{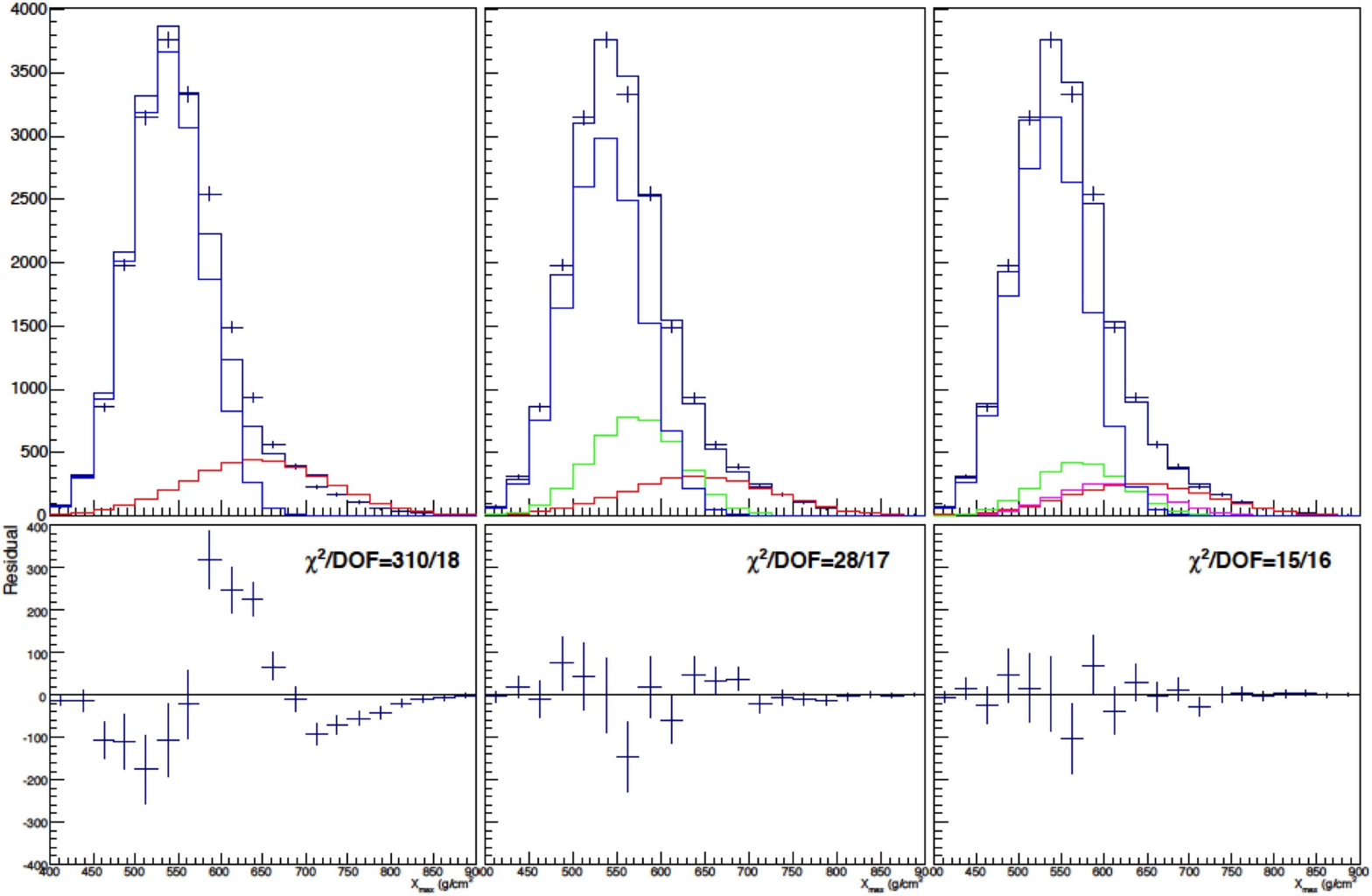}
  \caption{Fits to a simulated HECR composition at $10^{16.5}$ eV. 20,000 events (2 years data) were selected from a distribution of composed of H, He, CNO and Fe in the ratios 1:1:1:7. The individual components were then used as templates to fit the fraction of each. In the left panel, only H and Fe templates were used, leading to the large $\chi^2$ value for the fit and large residuals. The center panel shows the result of a three-component fit using H, CNO and Fe. 
The residuals are reduced, but the $\chi^2$ value is still relatively large. 
The right panel shows the four-component fit and demonstrates a significant improvement in both the residuals and the $\chi^2$. The four-component fit reconstructs the input composition ratio to within 10\% of the 1:1:1:7 input composition ratio. Note that the ordinate in each plot is \xmax\, in units of g/cm$^2$.}
\end{figure}



\bibliographystyle{aipproc}   % if natbib is available

% \bibliography{sample}

% \IfFileExists{\jobname.bbl}{}
%  \typeout{******************************************}
% \typeout{** Please run "bibtex \jobname" to optain}
%  \typeout{** the bibliography and then re-run LaTeX}
%  \typeout{** twice to fix the references!}
% \typeout{******************************************}

%\end{document}

%%%%%%%%%%%%%%%%%%%%%%%%%%%%%%%%%%%%%%%%%%%
%% The following lines show an example how to produce a bibliography
%% without the help of the BibTeX program. This could be used instead
%% of the above.
%%%%%%%%%%%%%%%%%%%%%%%%%%%%%%%%%%%%%%%%%%%

\end{document}